\documentclass{aa}
\usepackage[breaklinks,colorlinks,linkcolor=blue,citecolor=blue,urlcolor=blue,hyperfootnotes=true]{hyperref}
\usepackage{graphicx}
\usepackage{txfonts}
\usepackage{url}
\usepackage{graphicx}

\usepackage{xcolor}
\usepackage[normalem]{ulem}

\def\kms {km~s$^{-1}$}

\def\uas {$\mu$as}
\def\deg {\ifmmode {^\circ}\else {$^\circ$}\fi}
\def\porm {\ifmmode {\pm}\else {$\pm$}\fi}
\def\chisqpdf {\ifmmode {\chi^2_{\rm pdf}}\else {$\chi^2_{\rm pdf}$}\fi}
\def\chisq {\ifmmode {\chi^2}\else {$\chi^2$}\fi}

\def\etal {et al.~}

\def\d {\ifmmode {{\rlap{.}}^\circ}\else {${\rlap{.}}^\circ$}\fi}
\def\s {\ifmmode {{\rlap{.}}^s}\else {${\rlap{.}}^s$}\fi}
\def\as {\ifmmode {{\rlap{.}}^{''}}\else {${\rlap{.}}^{''}$}\fi}
\def\pa {\ifmmode {\psi} \else {$\psi$}\fi}
\def\vlsr {\ifmmode {v_{\rm LSR}}\else {$v_{\rm LSR}$}\fi}
\def\vlsrr {\ifmmode {v^r_{\rm LSR}}\else {$v^r_{\rm LSR}$}\fi}
\def\vhelio {\ifmmode {v_{Helio}}\else {$v_{Helio}$}\fi}
\def\To {\ifmmode{\Theta_0}\else{$\Theta_0$}\fi}
\def\Ro {\ifmmode{R_0}\else{$R_0$}\fi}
\def\Vlsr {\ifmmode {V_{\rm LSR}} \else {$V_{\rm LSR}$} \fi}
\newcommand {\HII}{\mbox{H\,\textsc{ii}}}
\newcommand {\HI}{\mbox{H\,\textsc{i}}}
\definecolor {malachite}{rgb}{0.34, 0.7, 0.22}

\begin{document}

\title{Local spiral structure based on the \textit{Gaia} EDR3 parallaxes}

\author{Y. Xu\inst{1,2}, L. G. Hou\inst{3,5}, S. B. Bian\inst{1,2}, C. J. Hao\inst{1,2}, D. J. Liu\inst{1,4}, J. J. Li\inst{1,2}, Y. J. Li\inst{1}
}

\institute{Purple Mountain Observatory, Chinese Academy of Sciences, Nanjing 210023, PR China 
\email{xuye@pmo.ac.cn}
\and
School of Astronomy and Space Science, University of Science and Technology of China, Hefei 230026, PR China
\and
National Astronomical Observatories, Chinese Academy of Sciences, 20A Datun Road, Chaoyang District, Beijing 100101, PR China
\and
College of Science, China Three Gorges University, Yichang 443002, PR China
\and
CAS Key Laboratory of FAST, National Astronomical Observatories, Chinese Academy of Sciences, Beijing 100101, PR China
}

\date{Accepted}
\titlerunning{Local Spiral Structure}
\authorrunning{Xu et al}

\abstract
{The astrometric satellite \textit{Gaia} is expected to significantly
  increase our knowledge as to the properties of the Milky Way. The
  \textit{Gaia} Early Data Release 3 (\textit{Gaia} EDR3) provides the
  most precise parallaxes for many OB stars, which can be used to
  delineate the Galactic spiral structure.}
{We investigate the local spiral structure with the largest sample of
  spectroscopically confirmed young OB stars available to
  date, and we compare it with what was traced by the parallax
  measurements of masers.}
%
{A sample consisting of three different groups of massive young stars,
  including O--B2 stars, O--B0 stars and O-type stars with parallax accuracies
  better than 10\% was compiled and used in our analysis.}
%
{The local spiral structures in all four Galactic quadrants within
  $\approx$5 kpc of the Sun are clearly delineated in detail. The
  revealed Galactic spiral pattern outlines a clear sketch of nearby
  spiral arms, especially in the third and fourth quadrants where the
  maser parallax data are still absent.
  These O-type stars densify and extend the spiral structure
  constructed by using the Very Long Baseline Interferometry
  (VLBI) maser data alone. The clumped
  distribution of O-type stars also indicates that the Galaxy spiral
  structure is inhomogeneous.}
{}
\keywords{ astrometry -- Galaxy: structure -- stars: early type --
  stars: masive -- masers }

\maketitle
%

\section{Introduction}

Mapping the Galactic spiral structure has long been a difficult issue
in astronomy, because the Sun is deeply embedded in the Galactic plane,
resulting in the superposition of multiple structures along the
observed line-of-sight. The copious dust extinction makes the
situation even worse.
In the 1950s, substantial progress in regards to tracing the spiral arm
segments in the solar neighborhood was first made by
\citet{Mor:52,Mor:53} with optically selected high-mass stars (OB
stars). Later on, the early $\HI$ surveys in the radio band extended
the studies of spiral arms to almost the entire Galactic
disk~\citep{ch52,vmo54,khc57,west57,Oort:58,Bok:59}. However, it was
soon realized that the $\HI$ results were not reliable due to the
influence of the noncircular motions and the kinematic
distance ambiguities. In the 1970s, \citet{gg76} proposed a famous
model of four-arm segments by using $\sim$100 optical- and radio-selected
$\HII$ regions, which was further explored by many researchers by an
integrated approach of mass data of various spiral tracers, including
high-mass young stars, $\HII$ regions, giant molecular clouds,
$\HI$ clouds, etc.\citep[e.g.,][]{rus03,hh14}. Until now, many
approaches have been pursued to decipher the morphology of the Galaxy
spiral arms \citep[e.g., see][for reviews]{Xu18a,sz20}. There is a
general consensus that a global spiral pattern exists in the Galactic
disk between Galactocentric distance of about 3 and 10~kpc. However,
considerable disagreements remain in regards to the finer details. The arm geometries,
orientations, and even the number of arms are still debated.

Measuring the distances as accurately as possible for many spiral
tracers would be the key to settle the disputes. Through measuring the
trigonometric parallax of masers in high-mass star formation regions,
the VLBI has achieved a
revolutionary breakthrough in the spiral structure of the Milky Way
\citep[][]{xrzm06}. The VLBI can now obtain parallax accuracies down
to a few \uas\ \citep[e.g.,][]{Sanna17}.
Accurate trigonometric parallaxes for about 200 high-mass star
formation regions have been measured. They are spread over about 
one-third of the entire Galactic disk, which is the subject of an updated view of
the Galactic spiral structure that has recently been proposed by
\citet{Reid:19}.
However, there is still a lack of observational data for many Galaxy
areas, especially in the fourth Galactic quadrant. In order to accurately
trace the extension of nearby spiral arm segments, the parallax
measurements of masers need to be supplemented.

By taking advantage of the second data release of the {\it Gaia} mission,
\citet[][Hereafter Paper I]{Xu18b} extended the nearby spiral arms
from the first and second Galactic quadrants to the third and fourth
quadrants. The spiral structure within $\approx$3~kpc of the Sun was
well depicted with massive, young stars (OB stars). To accurately
extend the nearby arm segments to more distant regions
($\gtrsim$3~kpc), higher accuracies of parallax data are crucial.
Recently, the {\it Gaia} mission released part of the third data set (the
{\it Gaia} Early Data Release 3, hereafter {\it Gaia} EDR3). The astrometric
data \citep{gaia2016,gaia2020} have been updated significantly, up to
a parallax accuracy of 20--30 $\mu$as, which enables us to reveal the
detailed local spiral structure in a wider range ($\approx$5 kpc).

\begin{table*}
\caption{Parallaxes and proper motions of OB stars from {\it Gaia}
  EDR3.}
\label{tab:ob} 
\centering
\begin{tabular}{ccccccccc}
\hline \hline
Name   & \textit{Gaia} EDR3 ID & RA   & Dec   &  $\pi$  & $\mu_x$  & $\mu_y$  & Spectral \\
&   & ($^{\circ}$) & ($^{\circ}$) & (mas)  & (mas yr$^{-1}$) & (mas yr$^{-1}$)  &  type   \\
(1)  & (2)   & (3)   & (4)   & (5)  & (6)   & (7)   & (8)     \\
\hline
\\
LS I +66 3 & 528563384392653312 & 0.1219  & 67.2168  & 0.948$\pm$0.012  & $-$1.80$\pm$0.01  & $-$2.43$\pm$0.01  & OB \\
$\left[\rm B55b \right]$ 5542   & 429950454253663360 & 0.1471  & 62.5549  & 0.413$\pm$0.023  & $-$3.55$\pm$0.02  & $-$1.17$\pm$0.03  & B2 \\
$\left[\rm B53 \right]$ 368     & 423149081478526464 & 0.2867  & 58.9799  & 0.261$\pm$0.015  & $-$3.32$\pm$0.01  & $-$2.70$\pm$0.02  & B0 \\
LS I +64 9  & 431771623466692608 & 0.3067  & 64.5883  & 0.332$\pm$0.015  & $-$2.76$\pm$0.01  & $-$0.83$\pm$0.01  & B2V \\
$\left[\rm B53 \right]$ 369     & 423157774492279680 & 0.3113  & 59.1428  & 0.264$\pm$0.014  & $-$3.02$\pm$0.01  & $-$1.50$\pm$0.01  & B2 \\
...\\
\hline
\end{tabular}
\tablefoot{Column (1) is the star name given in \citet{Skiff+2014};
  Column (2) lists the \textit{Gaia} EDR3 ID of the matched source;
  Columns (3) and (4) are the right ascension (RA) and declination
  (Dec) given by {\it Gaia} EDR3, and the reference epoch is J2016.0;
  Column (5) is the {\it Gaia} parallax and its 1$\sigma$ uncertainty;
  Columns (6) and (7) list the proper motions in the eastward
  ($\mu_{x} = \mu_{\alpha} \cos \delta$) and northward ($\mu_{y} =
  \mu_{\delta}$) directions, respectively, also given are their
  1$\sigma$ uncertainties; Column (8) is the spectral type derived
  from \citet[][]{Skiff+2014}. The full table is available at the
  CDS. }
\end{table*}

\begin{figure*}
\centering
\includegraphics[scale=0.29]{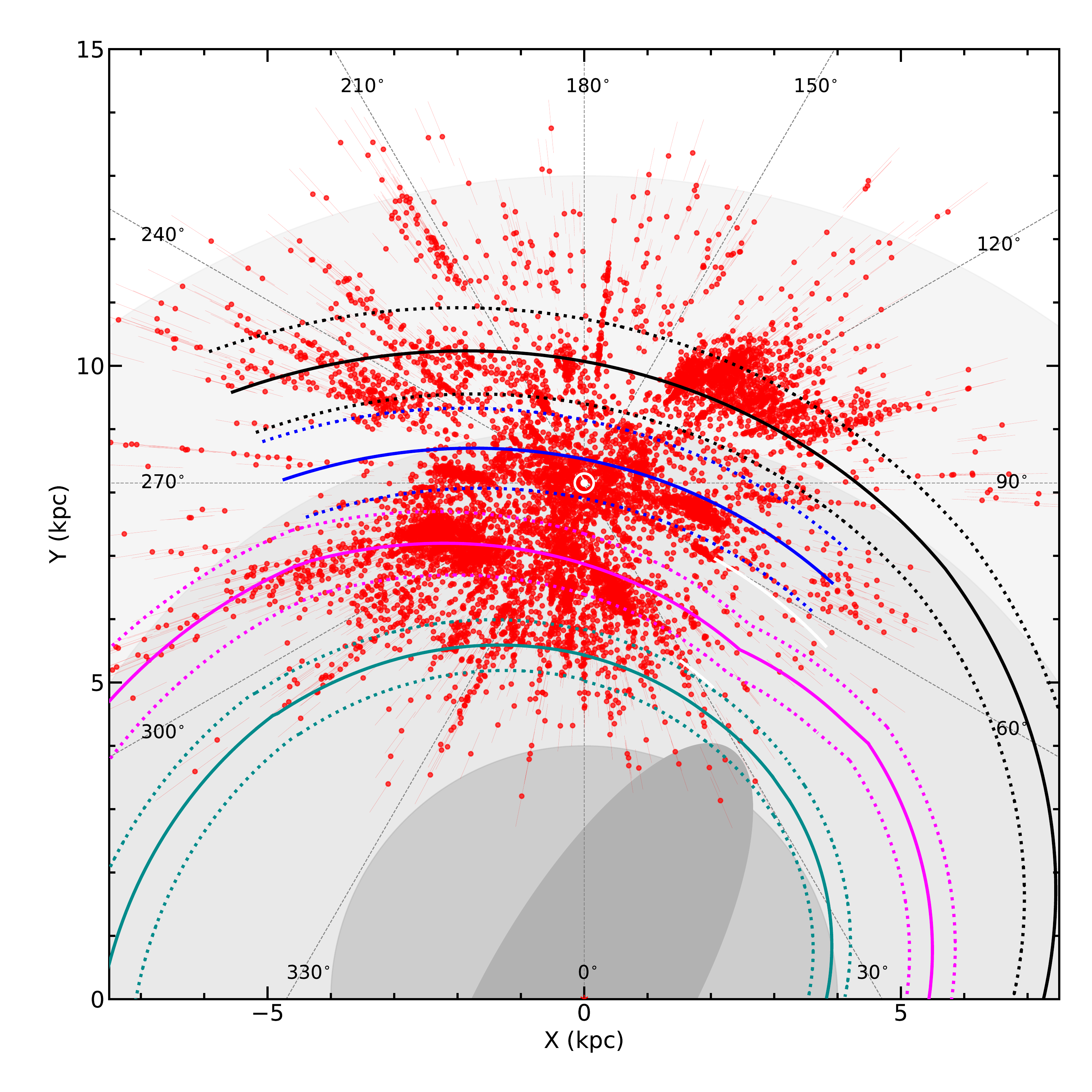}
\includegraphics[scale=0.29]{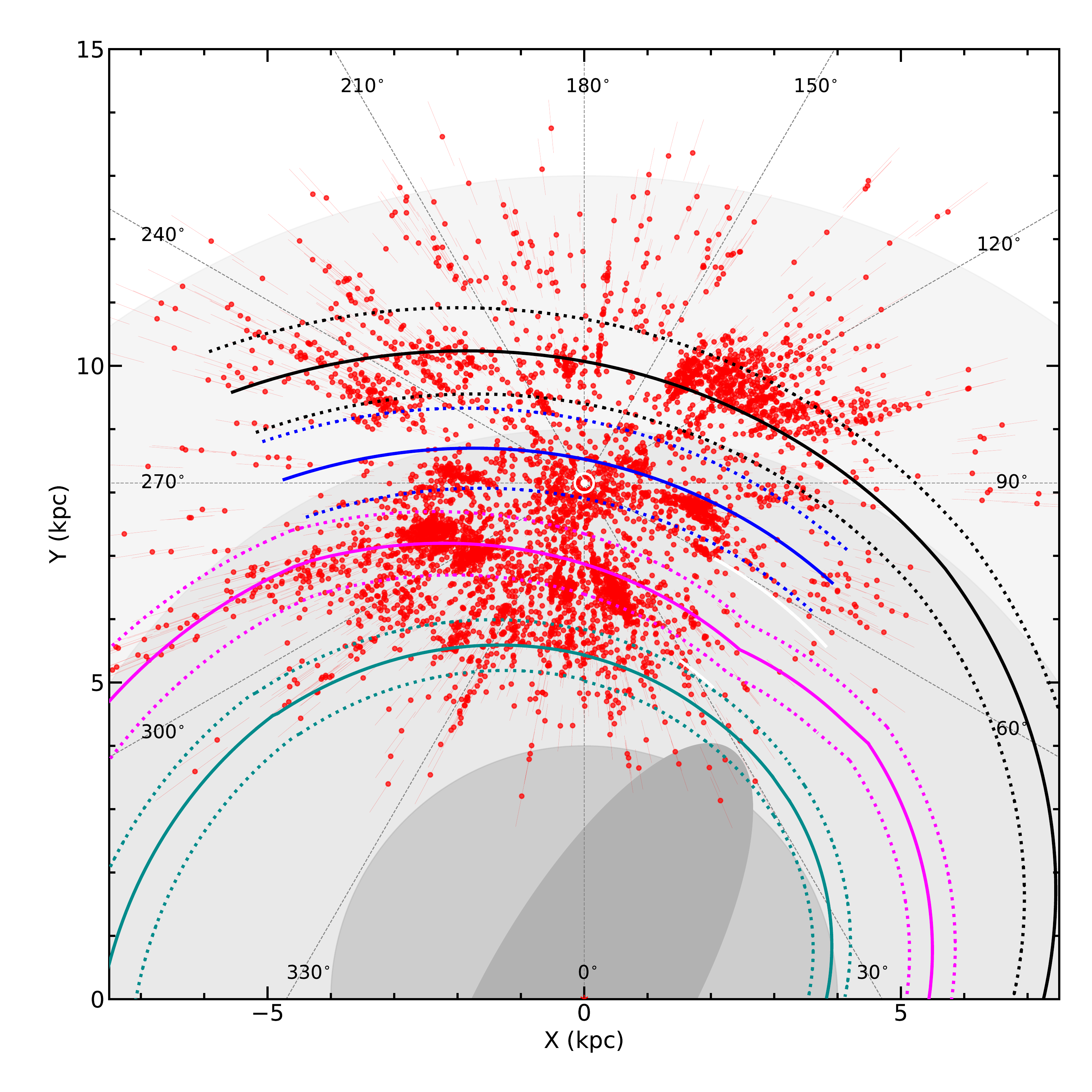}
\caption{Locations of the O--B2 stars (red dots, {\it Left}) and O--B0
  stars (red dots, {\it Right}) projected onto the Galactic plane,
  together with their 1$\sigma$ distance uncertainties.  The Sun (Sun
  symbol) is at (0, 8.15)~kpc. Only the stars with formal parallax
  uncertainties better than 10\% are shown. The solid and dashed
  curved lines denote the arm center and widths (enclose 90\% of the
  masers) fitted by \citet[]{Reid:19} from their parallax data of VLBI
  masers. The Perseus Arm (black), the Local Arm (blue), the
  Sagittarius-Carina Arm (magenta), the Scutum-Centaurus Arm (dark
  cyan), and two spur-like structures (white) are indicated with
  different colors. Straight dashed lines (gray) denote the Galactic
  longitudes.}
\label{fig:fig1}
\end{figure*}

\section{OB star sample}
\label{sec:data}
The OB stars are taken from the catalog of stellar spectral
classifications of \citet{Skiff+2014}, which has been updated up to 2020
February 6. This catalog represents 68612 spectroscopically
confirmed OB stars collected from the literature, along with
spectral type classifications and accurate coordinates ($\approx1
''$). It is probably the largest sample of spectroscopically
confirmed OB stars available to date, which enables us
to reveal the spiral structure within $\approx$5~kpc of the
Sun. However, this catalog is not complete, more OBs have yet
to be identified from {\it Gaia} or other photometric and spectroscopic
surveys.
We carried out a cross-match between these OB stars and the {\it Gaia}
EDR3 catalog with a match radius of 1$''$. After eliminating the
targets that had more than one positionally matched {\it Gaia} source,
we found 14414 O--B2 stars.

We noticed that many of the 14414 O--B2 stars have considerable
parallax uncertainties, that is to say larger than 10\%, which are
comparable with the typical spacing between the nearby spiral arm
segments.
Therefore, similar to Paper I, we only adopted the O--B2 stars with
distance accuracies better than 10\% to depict the nearby spiral
arms. After rejecting stars with vertical heights larger than
300~pc, we obtained a reduced subsample of 9750 sources that includes
5386 O--B0 stars and 1089 O-type stars. Their {\it Gaia} source ID,
coordinates, parallaxes, proper motions, and also the names and
spectral type classifications from \citet{Skiff+2014} are listed in
Table~\ref{tab:ob} \footnote{\url{https://cdsarc.u-strasbg.fr/ftp/vizier.submit//xygaia2020v2/}}. 
In comparison, there were only 2800 OB stars with
parallax accuracies better than 10\% in Paper I. For the OB stars
in Table~\ref{tab:ob}, their {\it Gaia} $G$ band apparent magnitudes
are between $\sim$2.6 and 18.3, $\sim$87\% of them have $G$
magnitudes $<12$. They are distributed in the Galactic longitude
range from 0$^{\circ}$ to 360$^{\circ}$, about 90\% of them are
located in Galactic latitude $|b|<$~5$^{\circ}$.

There is an average systematic bias of 0.017 mas on the parallax
zero-point of {\it Gaia} EDR3. How to properly correct a subsample
of {\it Gaia} stars for this bias is not a closed issue. We tested
its impact on our results by correcting the average systematic bias
to the stars, and we found that the influence is very small for the
nearby OB stars studied in this work. In the following analysis,
we neglect the bias on the {\it Gaia} parallax zero-point for
the OB stars.

\begin{figure*}
  \centering
  \includegraphics[scale=0.34]{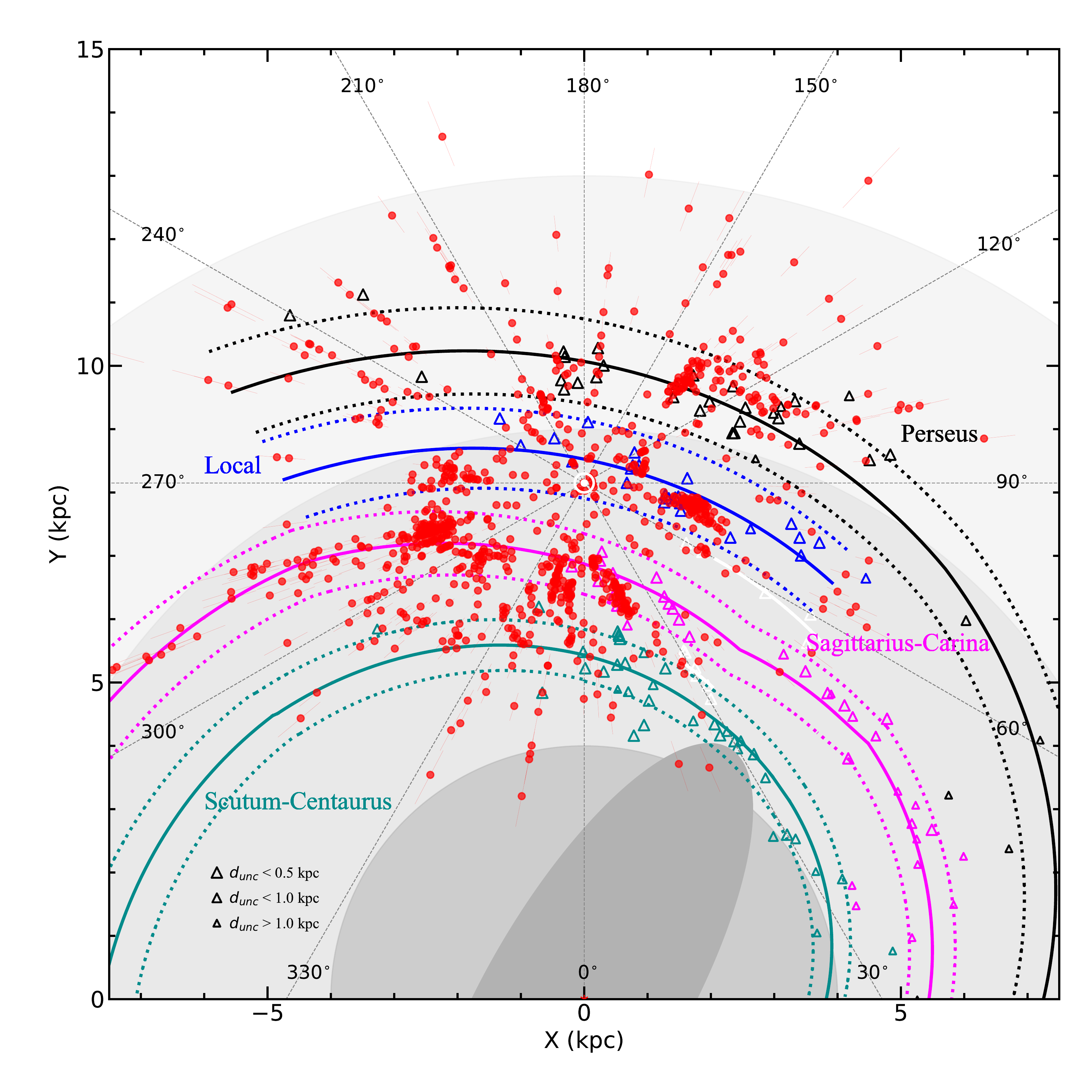}
  \caption{Locations of the O-type stars in {\it Gaia} EDR3 (red
    dots) and the masers (triangles) from \citet{Reid:19}. The
    formal parallax uncertainties of the O-type stars shown here are better
    than 10\%. See the caption of Fig.~\ref{fig:fig1} for more
    details.}
  \label{fig:fig2}
\end{figure*}

\section{Spiral structure}
\label{sect:Trac}
In this section, we investigate the nearby spiral structure traced by
three different groups of OB stars, that is, the O--B2 stars, the O--B0
stars, and only the O-type stars.
As shown in the left panel of Fig.\ref{fig:fig1}, the distribution of
O--B2 stars projected onto the Galactic plane clearly delineate three
spiral arm segments, they are the Perseus Arm, the Local Arm, and the
Sagittarius-Carina Arm from top to bottom.
In Paper I with {\it Gaia} DR2 data, the scopes of these arm segments
were only limited to within $\approx${3} kpc of the Sun. While with
the recently released {\it Gaia} EDR3 data, these arm segments extend
to more distant Galaxy regions, $\approx${5}~kpc away from the Sun as shown
in Fig.\ref{fig:fig1}. Especially in the fourth Galactic quadrant,
the Sagittarius-Carina Arm was traced to ($X\sim-7.0$, $Y\sim5.5$)~kpc
by O--B2 stars, as far as 7.5~kpc from us.
In the spiral arms, we notice that the distribution of O--B2 stars is
uneven, which may may be a result of the uneven distribution of giant
molecular clouds \citep[e.g.,][]{Dame:01, Sun:20} and/or
foreground extinction. Meanwhile, besides the major spiral arm
segments, there are quite a lot of stars scattered in the inter-arm
regions.

A peculiar motion speed of a star is normally from a few to tens of
\kms. The B2-type stars can live for $\sim$20 million years, which
means that they could migrate far from their birthplaces, up to about
1~kpc.
If their peculiar motions are perpendicular to spiral arms, a rather large
portion of the B2-type stars born in spiral arms would already have
migrated far from the birth sites.
In comparison, the lifetimes of B0-type stars are less than about ten
million years, only some of them could leave the spiral arms where they
were born. As shown in the right panel of Fig.\ref{fig:fig1}, relatively few stars are
located in the inter arm regions while we made a
similar plot with O--B0 stars, rather than O--B2 stars.

The more massive a star is, the younger it is, and the better tracer it is for
the Galaxy spiral arms.
Because of their shorter life spans than the B0--B2 stars, generally
speaking, most of the O-type stars are still located near their
birthplaces and, consequently, can be a better tracer for Galaxy spiral
structure.
In Fig.\ref{fig:fig2}, we solely focus on the distribution of O-type
stars. In order to depict the entire pattern of spiral arms, the
high-mass star formation region masers with VLBI parallax
measurements~\citep{Reid:19} are also displayed, which distribute in a
much broader region than that of O-type stars in the first Galactic
quadrant. By combining these high-quality data, the properties of
nearby spiral arms are explored and discussed below.

\begin{figure*}
  \centering
  \includegraphics[scale=0.26]{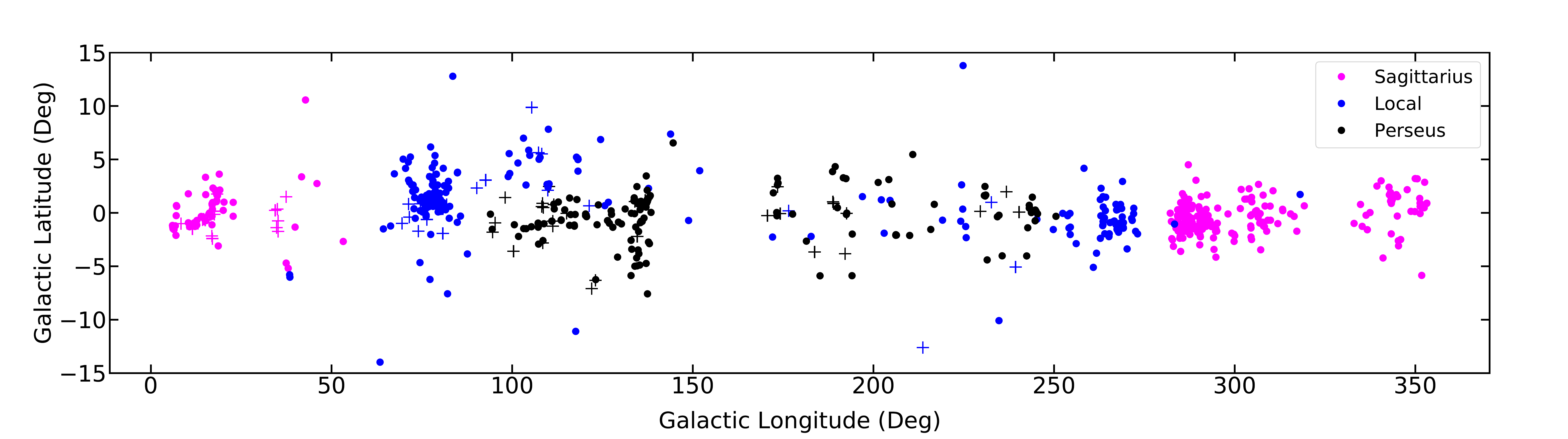}
  \caption{Galactic distribution of the O-type stars from {\it Gaia}
    EDR3 (dots) and the VLBI masers \citep[crosses,][]{Reid:19}. Only
    the O-type stars with parallax uncertainties $<$10\% are
    shown. Sources located in the Perseus Arm (black), the Local Arm
    (blue), and the Sagittarius-Carina Arm (magenta) are indicated by
    different colors.}
  \label{fig:fig3}
\end{figure*}

Along the Perseus Arm, the O-type stars tend to cluster.
Many O-type stars accumulate between Galactic longitude $l \sim$
{210}$^{\circ}$ and $l \sim$ {250}$^{\circ}$,
despite it being thought that a gap was there.
Together with clumps of O-type stars around $l \sim$ {180}$^{\circ}$
and clumps located between $l \sim$ {100}$^{\circ}$ and $l \sim$
{150}$^{\circ}$, these results indicate the uneven star formation in
the Perseus Arm.
Interestingly, in the third quadrant, the Perseus Arm traced by O-type
stars tends to spiral inward toward the Galactic center direction in
comparison to that defined by a small number of known VLBI masers.
In addition to pointing away and being traceable between heliocentric distances of
0.3 and 1.5 kpc, the Orion Spur protrudes from this major feature
between $l \sim$ 210$^{\circ}$ and $l \sim$ 220$^{\circ}$.

Remarkably, the Local Arm traced by the distribution of O-type stars
is distinct, it extends much longer than previously expected, and
it seems more similar to a major spiral arm feature.
It extends to the third and fourth Galactic quadrants
almost directly at $l \sim$ {260/270}$^{\circ}$, with a length of
$\sim$3~kpc from the Sun, and perhaps it would extend further and
spiral inward to the fourth quadrant.
In addition, there is an obvious difference between the distribution
of O-type stars and the modeled arm extension by using VLBI
masers. The Local Arm traced by O-type stars seems to present a
downward drift toward the Galactic center in comparison of the maser
measurements. In the longitude range of $l\sim240^\deg-330^\deg$, there
is still a lack of VLBI maser data, the upcoming BeSSeL Survey to the
southern hemisphere would solve the mystery \citep{Reid:19}.

The Sagittarius-Carina Arm delineated by the distribution of O-type
stars is shown as an arc-shaped structure: from $l\approx 50^{\circ}$
through $l = 0^{\circ}$ to $l\approx -80^{\circ}$, between $R$
$\sim$6~kpc and $\sim$9~kpc with a length of $\approx${10}~kpc, where
$R$ is the Galactocentric radius.
Strikingly, in the fourth quadrant, the distribution of O-type stars
in the Sagittarius-Carina Arm is well consistent with the model
prediction of arm extension given by the VLBI masers
\citep{Reid:19}.
Meanwhile, the Scutum-Centaurus Arm is also discernable from $l\approx
25^{\circ}$ through $l = 0^{\circ}$ to $l\approx -50^{\circ}$ in the
distribution of O-type stars.

In short, the O-type stars tend to clump together and some of them are
frequently found between the major spiral arms, indicating that recent
star formation does occur in the inter arm regions and/or some of them
have migrated far away from their birthplaces.
The nearby spiral pattern, on the whole, is similar from both VLBI and
{\it Gaia} parallax measurements in the range of $0^{\circ} <$
  {\it l} $< 240^{\circ}$.
Meanwhile, as shown in Fig.\ref{fig:fig3}, the distributions of O-type
stars in the Galactic latitude almost resemble the VLBI masers,
indicating that the O-type stars indeed trace the spiral arms similar
to the masers. In Fig.\ref{fig:fig3}, to assign the O-type stars
to the nearby spiral arms, we adopted the best-fitted arm center and arm widths
(enclose 90\% of the VLBI maser sources) from \citet{Reid:19}.

\section{Conclusions}
The VLBI parallax measurements of masers have nicely traced the Galactic
spiral structure in the first, second, and third Galactic
quadrants. While there is still a lack of maser data in about two-thirds
of the entire Galactic disk, especially for the fourth
quadrant. The OB stars with accurate {\it Gaia} parallaxes can densify
and extend the current picture, in particular in the Galaxy area
within $\approx$5~kpc of the Sun. In this work, with the compiled
O-type stars, the Perseus Arm, the Local Arm, the Sagittarius-Carina
Arm, and the Scutum-Centaurus Arm are extended to the third or even
fourth Galactic quadrants with high confidence, where the maser data
are still rare.
The O-type stars tend to clump together, largely as a result of the
state of their ancestors, the giant molecular clouds, indicating that
the Milky Way presents a discrete or uneven spiral structure.
The ubiquitousness of spurs traced by O-type stars also indicate that our
Galaxy may not have a pure grand design morphology.

\begin{acknowledgements}
This work was sponsored by the MOST under grant No. 2017YFA0402701. 
This work was funded by the NSFC Grands 11933011, 11873019 and
11673066, 11988101 and the Key Laboratory for Radio Astronomy. L.G.H
thanks the support from the Youth Innovation Promotion Association
CAS. This work has made use of data from the European Space Agency
(ESA) mission {\it Gaia} (\url{https://www.cosmos.esa.int/gaia}),
processed by the {\it Gaia} Data Processing and Analysis Consortium
(DPAC,
\url{https://www.cosmos.esa.int/web/gaia/dpac/consortium}). Funding
for the DPAC has been provided by national institutions, in particular
the institutions participating in the {\it Gaia} Multilateral
Agreement.
\end{acknowledgements}


%
%


\end{document}